\documentstyle[prd,aps,preprint,epsf]{revtex}
\begin{document}
\tightenlines
\draft
\widetext

\title{\Large The wave function of the universe and spontaneaus breaking
of supersymmetry  }
\author{O. Obreg\'on\thanks{E-mail: octavio@ifug3.ugto.mx},
J.J. Rosales\thanks{E-mail: juan@ifug3.ugto.mx} 
 J. Socorro\thanks{E-mail: socorro@ifug4.ugto.mx} 
and V.I. Tkach\thanks{E-mail: vladimir@ifug1.ugto.mx} \\
Instituto de F\'{\i}sica de la Universidad de Guanajuato,\\
Apartado Postal E-143, C.P. 37150, Le\'on, Guanajuato, Mexico.
}

\date{\today}

\maketitle

\begin{abstract}
In this work we define a scalar product ``weighted'' with 
the scalar factor $R$ and
show how to find a normalized wave function for the  supersymmetric 
quantum FRW cosmological model  using the idea of supersymmetry 
breaking selection rules under local n=2 conformal supersymmetry. We also
calculate the expectation value of the scalar factor R in this model and 
its corresponding behaviour. 

\end{abstract}
\vspace{0.5cm}
PACS numbers: 11.30.Pb, 12.60.Jv, 98.80.Hw
\narrowtext
\newpage

\section {Introduction}
Our universe as a whole can hardly be called a usual quantum system, 
however, if we want to apply the laws of quantum mechanics to such 
exiting processes as the birth of the universe and its subsequent evolution 
we have  to treat it as a quantum object obeying some fundamental
principles of the usual quantum mechanics including unitarity, and hence the
normalizability of the wave function of the universe and other
characteristics that only appear in the quantum level.
In the usual Hamiltonian formulation the wave function of the universe 
$\Psi(g_{ij},\phi)$ is a functional only of the metric field $g_{ij}(x^i)$ 
and other fields $\phi(x^i)$ defined on three-space, so
the relation in classical general relativity $H\Psi=0$, known as the
Wheeler-DeWitt (WD) equation, seems to indicate that theories no causality in
quantum cosmology, no dynamics and no obvious analogue of a conserved
probability distribution.
Quantum cosmology has revived a very old issue in quantum theory, i.e. the 
relation between wave function and probability distributions \cite{HP},
but it is well known, that the Hartle-Hawking wave function is not 
normalizable in a semiclassical approximation\cite{HP,HH}, and also it does
not allow a conserved current with a positive-definite probability density.  
A partial answer to this subject  was introduced, for first
time in \cite{Ma}, where a ``weight'' function is used in the definition of a
positive-definite
probability density $\rho$, using the Dirac quantization for FRW cosmologies. 
Here we will give other answer to this subject, using supersymmetric 
quantum approach 
n=2, where we also introduce one ``weight'' function in the definition for
the inner product of two supersymmetry states, that permits us to avoid
the ordinary  real parameter ``p'' that measure the ambiguity in the factor 
ordering (here is fixed!). 

 Other intrisic problem is selection of the appropriate  
boundary conditions on the wave function. To solve this problem,
some proposals have been suggested \cite{HH,Vi,L,Mena,BK}. 
Also at the quantum level the problem of the ambiguity in the factor ordering,
when the WD equation contains the 
product of non-commuting operators $\xi$ and 
$\pi_\xi$, where $\xi$ is any canonical ``coordinate'' field and $\pi_\xi$ is
the canonical conjugate momentum to this ``coordinate''. In the usual way, this
ambiguity has been ``measured'' with the parameter ``p''\cite{HH,Gri}, where p 
is a real constant. For some particular ``p'' values it was possible
to solve the WD equation \cite{OS,RS}. 

In this work we present an approach to  solve this last problem, using the 
ideas of supersymmetry breaking selection rules under local n=2 conformal 
supersymmetry. We present  a solution for the wave function of the universe
for the FRW cosmological model,
that is normalizable, and also we obtain
the behaviour of the spectation value for the scalar factor in the usual way,
giving us proportional to $\ell_{pl}^{2/3}$. 
The existence of normalizable solution of the system, that appears in this 
approach means in its turn, that supersymmetry is unbroken quantum 
mechanically.

To solve this problem in our 
supersymmetric approach the ``coordinate'' field corresponds to the scalar 
factor of the universe for the FRW cosmological model, when we do the 
integration with 
measure $R^{1/2} dR$ in the inner product of two states, thus the momenta 
hermitian-conjugate to $\pi_R$ is non-hermitian with 
$\pi_R^\dagger=R^{-1/2} \pi_R R^{1/2}$. However, we can construct one hermitian
operator and by means of this  avoid the use of the real parameter ``p''.  

This work is organized in the following way. In Sec. II we present the 
lagrangian, including the auxiliary fields and with them we give the
supersymmetry breaking selection rules. In Sec. III we give the 
super-algebra,
that satisfies the different operators in the theory. Sec. IV is devoted to
the solution of the normalized wave function state with 
zero energy, besides, we present the spectation value for the scalar factor.
Finally, Sec. V is for conclusions and remarks.

\section{Supersymmetric Lagrangian and Susy breaking}

The most general superfield action for a homogeneous scalar supermultiplet 
interacting with the scalar factor in the supersymmetric FRW model
\cite{ORT1,TRM} has the form
\begin{eqnarray} 
\rm S &=& \rm S_{FRW} + S_{\rm mat}, \nonumber \\
S_{FRW}&=& \int 6\left[- \frac{1}{2\kappa^2}\frac{{I\!\!R}}{{I\!\!N}} 
{\cal D}_{\bar \eta} {I\!\!R} {\cal D}_\eta {I\!\!R}
+\frac{\sqrt{k}}{2\kappa^2} {I\!\!R}^2 \right]
d\eta d {\bar \eta} dt\, , \nonumber\\            
 S_{ mat} &=&  \int \left[ \frac{1}{2} \frac{{I\!\!R}^3}{{I\!\!N}} 
{\cal D}_{\bar \eta} {\Phi} {\cal D}_\eta {\Phi}  
- 2 {I\!\!R}^3 g({\Phi} )\right] d\eta d {\bar \eta} dt\, ,  
\label{action1}
\end{eqnarray}
with $k=0,1$ stands for flat and closed space, and 
$\kappa^2 =8\pi G_N$, where $G_N$ is Newton's constant of gravity,
 $(\hbar =c=1)$. The units for the constants and fields in this
work are the following: $[\kappa^2]=\ell^2$, $[{I\!\!N}]=\ell^0 , 
[{I\!\!R}]=\ell^{1}, [{\Phi} ]=\ell^{-1}$  and
superpotential $[g({\Phi} )]=\ell^{-3}$, where $\ell $ correspond to units of 
lenght

For the one-dimensional gravity superfield ${I\!\!N} (t,\eta ,\bar\eta)$ 
$({I\!\!N}^\dagger ={I\!\!N})$ we have the following series expansion

\begin{equation}
{I\!\!N}(t,\eta ,\bar\eta )=N(t)+i\eta\bar\psi^\prime (t)+i\bar\eta\psi^\prime
(t)+\eta\bar\eta {\cal V}^\prime (t) \, ,
\label{expansion1}
\end{equation}
where $N(t)$ is the lapse function and also we have introduced the 
reparametrization $\psi^\prime (t)= N^{\frac{1}{2}} (t)\psi (t)$ 
and ${\cal V}^\prime (t) = {I\!\!N}(t) {\cal V}(t) +\bar \psi (t)\psi (t)$.

The Taylor series expansion for the superfield ${I\!\!R} (t,\eta ,\bar\eta)$
with the scalar factor $R(t)$ has the similar form

\begin{equation}
{I\!\!R} (t,\eta ,\bar\eta)= R(t)+i\eta\bar\lambda^\prime (t)
+i\bar\eta\lambda^\prime (t) +\eta\bar\eta {\cal B}^\prime (t),
\label{expansion2}
\end{equation}
where $\lambda^\prime (t) =\kappa N^{\frac{1}{2}} (t)\lambda (t)$ and
${\cal B}^\prime (t)=\kappa N(t){\cal B}(t)+\frac{\kappa}{2} 
(\bar\psi \lambda -\psi \bar\lambda)$.  

The component of the (scalar) matter superfields $\Phi (t,\eta ,\bar\eta)$
may be written as $(\Phi^+ =\Phi )$

\begin{equation}
\Phi =\varphi (t)+i\eta\bar\chi^\prime (t)+i\bar\eta \chi^\prime (t)+F^\prime
(t)\eta \bar \eta
\label{matter}
\end{equation}
where $\chi^\prime (t) =N^{\frac{1}{2}} (t)\chi (t)$ and $F^\prime (t)=NF+
\frac{1}{2} (\bar \psi \chi -\psi \bar \chi )$, 
besides,
${\cal D}_\eta = \frac{\partial}{\partial\eta}+i \bar\eta 
\frac{\partial}{\partial t}$ and ${\cal D}_{\bar\eta}=-
\frac{\partial}{\partial\bar\eta}-i\eta \frac{\partial}{\partial t}$ are the
supercovariant derivatives of the conformal supersymmetry $n=2$, which has
dimension $[{\cal D}_\eta ]=[{\cal D}_{\bar\eta}] =\ell^{-\frac{1}{2}}$. 

After the integration over Grassmann  complex coordinate $\eta$ and 
$\bar\eta$, and making the following redefinition of the ``fermion"  fields 
(Grassmann variables) 
$\lambda (t) \to \frac{1}{3} R^{-\frac{1}{2}}(t)\lambda(t)$ and 
$\chi (t) \to R^{-3/2} (t) \chi (t)$, we find the Lagrangian in which 
the fields ${\cal B}(t)$ and $F(t)$
are auxiliary and they can be eliminated with the help of their equations 
of motion.

Finally, the Lagrangian in terms of components of the superfields 
${I\!\!R}$, ${I\!\!N}$, $\Phi$ takes the form

\begin{eqnarray}
L &=&-\frac{3}{\kappa^2} \frac{R({\cal D} R)^2}{N} +\frac{2}{3} i\bar\lambda
{\cal D}\lambda +\frac{\sqrt{k}}{\kappa} R^{\frac{1}{2}} (\bar\psi \lambda -
\psi\bar\lambda ) \nonumber \\
&+& \frac{1}{3} NR^{-1} \sqrt{k} \bar\lambda\lambda +\frac{3k}{\kappa^2} NR+
\frac{R^3}{2}\frac{({\cal D}\varphi)^2}{N}-i\bar\chi {\cal D}\chi \nonumber \\
&-& \frac{3}{2} \sqrt{k}NR^{-1}\bar\chi\chi -\kappa^2 N g(\varphi )\bar\lambda
\lambda -6\sqrt{k} N g(\varphi)R^2  \label{lagra}  \\
&-&NR^3 V(\varphi)+\frac{3}{2}\kappa^2 N g(\varphi)\bar\chi\chi 
+\frac{i\kappa}{2} {\cal D}\varphi (\bar\lambda \chi +\lambda\bar\chi ) 
\nonumber \\
&-& 2 N \frac{\partial^2 g(\varphi)}{\partial\varphi^2} \bar\chi\chi 
-\kappa N \frac{\partial g(\varphi)}{\partial\varphi}(\bar\lambda\chi 
- \lambda\bar\chi ) + 
\frac{\kappa^2}{4}R^{-3/2}(\psi\bar\lambda -\bar\psi\lambda )\bar\chi\chi 
\nonumber \\
&-& \kappa R^{3/2} (\bar\psi\lambda - \psi\bar\lambda) g(\varphi)+R^{3/2} 
\frac{\partial g(\varphi)}{\partial \varphi} (\bar\psi \chi -\psi \bar \chi)\,
 ,  \nonumber
\end{eqnarray} 
where ${\cal D} R=\dot R-\frac{i\kappa}{6} R^{-\frac{1}{2}} 
(\psi\bar\lambda +\bar\psi\lambda )$ and 
${\cal D}\varphi =\dot \varphi -\frac{i}{2} R^{-\frac{3}{2}}
(\bar\psi\chi +\psi\bar\chi)$ are the supercovariant derivatives, and 
${\cal D}\lambda =\dot\lambda -\frac{i}{2} {\cal V}\lambda$, 
${\cal D}\chi =\dot\chi -\frac{i}{2}{\cal V}\chi$ are the ${\cal U}$ (1) 
covariant derivatives.

The potential for the homogeneous scalar fields 
\begin{equation}
V(\varphi ) =2 \left( \frac{\partial g(\varphi)}{\partial\varphi}\right)^2
- 3\kappa^2 g^2(\varphi),
\label{potential}
\end{equation}
consists of two terms, one of them is the potential for the scalar field in 
the case of global supersymmetry. The potential (\ref{potential}) 
 is not positive semi-definite in contrast with the standard 
supersymmetric quantum mechanics.
Unlike the standard supersymmetric quantum mechanics, our model,  
 describing the minisuperspace approach to
supergravity coupled to matter, allows the  supersymmetry breaking when the 
vacuum energy is equal to zero $V(\varphi)=0$.  

In order to give some implications of spontaneous supersymmetry breaking we 
display the potential $V(\varphi)$ (\ref{potential}) in terms of the 
auxiliary fields $F(t)$ and ${\cal B}(t)$, 

\begin{equation}
V(\varphi) = \frac{1}{2} F^2 - \frac{3}{\kappa^2 R^2} {\cal B}^2\, ,
\label{po-aux} 
\end{equation}
where the bosonic $F$ and ${\cal B}$  are
\begin{equation}
F=2 \frac{\partial g(\varphi)}{\partial\varphi } \, , \,\qquad 
{\cal B}= -\kappa^2 R\, g(\varphi) \, .
\end{equation}

The selection rules for the occurrence of spontaneous supersymmetry breaking 
are

\begin{eqnarray}
&{\rm i)}&\qquad \frac{\partial V(\varphi)}{\partial\varphi} 
= 4\frac{\partial g}{\partial\varphi} 
\left[ \frac{\partial^2 g}{\partial\varphi^2} 
-\frac{3}{2} \kappa^2 g \right]=0, 
\,\, {\rm at} \,\, \varphi = \varphi_0 \label{uno}\\
&{\rm ii)}&\qquad V (\varphi_0 )= 0 \Rightarrow
 \left[ \left( \frac{\partial g}{\partial\varphi} \right)^2 -\frac{3}{2} 
\kappa^2 g^2 \right] =0 , \label{dos}\\
&{\rm iii)}&\qquad F= 2\frac{\partial g(\varphi)}{\partial\varphi} \not =
0 \, , \, {\rm at} \, \varphi = \varphi_0, \label{tres} 
\end{eqnarray}

The first condition implies the existence of a minimum, in the scalar field;
the second condition  is the absence of the cosmological constant, the third 
condition  is for the  breaking of supersymmetry.
The measure of this breakdown is the term $-\kappa^2 g(\varphi )
N\bar\lambda\lambda$ in the lagrangian (\ref{lagra}). Furthermore, we can 
identify  
\begin{equation}
m_{3/2} = \kappa^2 g(\varphi_0)= \frac{g(\varphi_0)}{M^2_{pl}}
\label{mass}
\end{equation}
as the gravitino mass in the effective supergravity theory, and 
$M_{pl}=\frac{1}{\kappa} = \frac{1}{\sqrt{ 8\pi G_N}}= 2.4 \times 10^{18}Gev$
is the reduced Planck mass.
 
The factor $R$ in the kinetic terms of the  scalar factor 
$-\frac{3}{\kappa^2N} R (\dot R)^2$ plays the role of a ``metric'' tensor in 
the Lagrangian, the kinetic energy for the scalar factor is negative 
definite, this is due to the fact, that the particle-like fluctuations do not 
correspond to the scalar factor, and the kinetic energy of the scalar 
fields $\frac{R^3}{2N} \dot\varphi^2$ 
is positive.
\section{The corresponding supersymmetric quantum mechanics}

The hamiltonian can be calculated in the usual way. We have the classical 
canonical Hamiltonian

\begin{equation}
H_{can} =NH+\frac{1}{2}\bar\psi \, S-\frac{1}{2}\psi \, \bar S+ 
\frac{1}{2} {\cal V} \, {\cal F} ,
\label{hamiltonian}
\end{equation} 
where $H$ is the Hamiltonian of the system, $S$, $\bar S$  are supercharges and
${\cal F}$ is the  ${\cal U}(1)$ rotation generator. The form of the canonical
Hamiltonian (\ref{hamiltonian}) explains the fact, that $N, \psi, \bar\psi$ 
and ${\cal V}$ are  Lagrange multipliers which enforce  only the first class 
constraints $H=0, S=0, \bar S=0$ and ${\cal F}=0$, which express the 
invariance of the conformal
$n=2$ supersymmetric transformations. As usual with the Grassmann variables,
we have the second-class constraints which can be eliminated by the Dirac 
procedure, as a result only the following non-zero 
Dirac brackets $\left\{\,\, ,\,\,\right\}$ remain. For Grassmann  variables 
$\lambda ,\bar\lambda ,\chi$
and $\bar\chi$ 

\begin{equation}
\left\{\lambda ,\bar\lambda \right\}_*=+\frac{3}{2}i \, , \, 
\left\{\chi ,\bar\chi \right\}_* =-i \, .
\label{bracket1}
\end{equation}

The canonical Poisson brackets for the $R, \pi_R$ and $\varphi, \pi_\varphi$,
 have the following form
\begin{equation}
\left\{R, \pi_R\right\}_{Pb} =-1 \, , \, \left\{\varphi ,\pi_\varphi 
\right\}_{Pb} =-1 \, .
\label{bracket2}
\end{equation}

In a quantum theory the brackets (\ref{bracket1}) and (\ref{bracket2}) must be 
replaced by anticommutator

\begin{equation}
\left\{\lambda , \bar\lambda \right\}=-\frac{3}{2} \, , \, 
\left \{\chi ,\bar\chi \right\} =1 ,
\label{bra}
\end{equation}
and can be considered as generators of the Clifford algebra,  as well as the 
commutators
\begin{equation}
\left[R,\pi_R\right] =-i \, , \,\left [\varphi , \pi_\varphi \right] =-i\, .
\end{equation}

The quantization procedure takes  into account the dependence of the 
Lagrangian on the metric ``R", but at the quantum level we must consider the
nature of the Grassmann variables $\lambda ,\bar\lambda ,\chi$ and $\bar\chi$,
and with respect to these ones we perform the antisymmetrizations, then
 we can write the bilinear combination in the form of the commutators e.g. 
$\bar\chi \chi \to \frac{1}{2}\left [\bar\chi ,\chi \right]$ and this leads 
to the 
following quantum 
Hamiltonian $H$
\begin{eqnarray}
H &=& -\frac{\kappa^2}{12} R^{-\frac{1}{2}} \pi_R R^{-\frac{1}{2}} 
\pi_R -
\frac{3k R}{\kappa^2} -\frac{1}{6} \frac{\sqrt{k}}{R} \left[\bar\lambda , 
\lambda \right]
+ \frac{\pi^2_\varphi}{2R^3}  \nonumber \\
&&- \frac{i\kappa}{4R^3}\pi_\varphi \left(\left[\bar\lambda ,\chi \right]
+\left[\lambda ,\bar\chi \right] \right)
-\frac{\kappa^2}{16R^3} \left[\bar\lambda ,\lambda \right] 
\left[\bar\chi ,\chi \right]+ \frac{3\sqrt{k}}{4 R} \left[\bar\chi , \chi 
\right] \nonumber \\
&&+ \frac{\kappa^2}{2} g(\varphi) \left[\bar\lambda ,\lambda \right]
+6\sqrt{k}\,  g(\varphi) R^2 + R^3 V(\varphi) +\frac{3}{4} \kappa^2 
g(\varphi) \left[\bar\chi ,\chi \right]  \nonumber \\
&& + \frac{\partial^2 g(\varphi)}{\partial \varphi^2} \left[\bar\chi ,\chi 
\right]+ \frac{1}{2}\kappa \frac{\partial g(\varphi)}{\partial\varphi} 
\left( \left[ \bar\lambda ,\chi \right]-\left[\lambda ,\bar\chi \right]
\right) \, ,
\label{hamil}
\end{eqnarray}
where $\pi_R = i \frac{\partial}{\partial R}$ and
$\pi_\varphi = i \frac{\partial}{\partial \varphi}$. 
The supercharges $S$, $S^\dagger$ and the fermion number operator ${\cal F}$ 
have the following structures

\begin{equation}
S=A\lambda +B\chi \, , \qquad \, S^\dagger =A^\dagger \lambda^\dagger 
+B^\dagger  \chi^\dagger \, ,
\end{equation}
 where  
\begin{eqnarray}
A&=&\frac{i}{3}\kappa R^{-\frac{1}{2}} \pi_R - \frac{2\sqrt k}{\kappa}
R^{\frac{1}{2}} + 2\kappa R^{\frac{3}{2}} g(\varphi) + \frac{\kappa}{4}
R^{-\frac{3}{2}} \left[ \bar \chi, \chi \right]\, , \nonumber \\
B&=& i R^{-\frac{3}{2}} \pi_\varphi + 2 R^{\frac{3}{2}} 
\frac{\partial g(\varphi)}{\partial \varphi} \, ,
\end{eqnarray}
with $A^\dagger$ and $B^\dagger$ are hermitian to A and B respectively, 
and 
\begin{equation}
{\cal F}=-\frac{1}{3}\left[\bar\lambda ,\lambda \right]
+\frac{1}{2} \left[\bar\chi ,\chi \right] .
\end{equation}

An ambiguity exist in the factor ordering of these operators, such 
ambiguities always arise, when the operator expression contains the product of
non-commuting operators $R$ and $\pi_R$ as in our case. It is then 
necessary to find some criteria to know which factor ordering should be 
selected. We propose the following: to integrate with measure 
$R^{\frac{1}{2}}dR$ in 
the inner product of two states \cite{AFF}. In this measure the 
conjugate momentum  $\pi_R$ is non-hermitan with 
$\pi^\dagger_R=R^{-\frac{1}{2}} \pi_R R^{\frac{1}{2}}$,
however, the combination $(R^{-\frac{1}{2}}\pi_R)^\dagger =\pi^\dagger_R 
R^{-\frac{1}{2}} = R^{-\frac{1}{2}} \pi_R$ is a hermitian one, and  
$(R^{-\frac{1}{2}} \pi_R R^{-\frac{1}{2}} \pi_R)^\dagger =
R^{-\frac{1}{2}} \pi_R R^{-\frac{1}{2}} \pi_R$ is 
hermitian too, with this the parameter $p=1/2$ is fixed.

The anticommutator value $\{\lambda , \bar\lambda \}=-\frac{3}{2}$ of the
superpartners $\lambda$ and $\bar\lambda$ of the scalar factor $R$ is negative,
unlike the anticommutation relation for $\chi$ and $\bar\chi$ in (\ref{bra}),
 which is positive. They can be redefined and one becomes the following 
conjugate operation for the operators $\lambda$ and $\chi$

\begin{equation}
\bar\lambda = \xi^{-1} \lambda^\dagger \xi =-\lambda^\dagger \, , \, \qquad
\bar\chi = \xi^{-1}
\chi^\dagger \xi = \chi^\dagger ,
\end{equation}
where $\{\lambda ,\lambda^\dagger \} =\frac{3}{2}$, and the operator $\xi$ 
possesses the following properties

\begin{equation}
\lambda^\dagger \xi =-\xi\lambda^\dagger \, , \,\qquad 
\chi^\dagger \xi=\xi \chi^\dagger \,\qquad {\rm and} \,\qquad
\xi^\dagger = \xi \, .
\end{equation}

So, for the supercharge operator $\bar S$ we have  the following equation

\begin{equation}
\bar S = \xi^{-1} S^\dagger \xi \, .
\end{equation}

For the quantum generators $H, S, \bar S$ and ${\cal F}$ we obtain the 
following superalgebra 

\begin{eqnarray}
\left\{ S, \bar S\right\} &=&2H \, ,\quad \, S^2=\bar S^2 =0 \, , \qquad \, 
\left[S,H\right]=0, \nonumber\\
\left[\bar S,H\right]&=&0 \, , \quad \left [{\cal F}, S\right]=-S \, ,\quad 
\,\left [{\cal F}, \bar S \right ]=\bar S \,  . 
\label{superalgebra}
\end{eqnarray}
We can see, that the anticommutator of supercharges $S$ and their conjugate
$\bar S$ under our conjugate operation has the form

\begin{equation}
\overline{ \left \{ S,\bar S \right\} }= \xi^{-1} 
\left\{ S,\bar S \right \}^\dagger 
\xi = \left\{S,\bar S\right\} , 
\end{equation}
and the  Hamiltonian operator is self-conjugate under the operation 
$\bar H =\xi^{-1} H^\dagger \xi$. In the case of standard supersymmetric 
quantum mechanics
we would have $\bar\lambda =\lambda^\dagger$ and $\bar S=S^\dagger$, and 
the Hamiltonian would be positive. In our case, the algebra 
(\ref{superalgebra}) 
does not define positive-definite Hamiltonian in a full agreement with the 
circumstance that the  potential $V(\varphi)$ (\ref{potential}) of the scalar 
field is not positive semi-definite in general, in contrast with the 
standard supersymmetric quantum mechanics.

We can choose the following matrix representation for the operators 
$\lambda ,\bar\lambda , \chi , \bar\chi$ and $\xi$ in the form of the 
tensorial product of matrices $2\times 2$
\begin{eqnarray}
\lambda &=&\sqrt{\frac{3}{2}}\sigma_{(-)}\otimes 1 \, 
,\qquad \,\bar\lambda =-\sqrt{\frac{3}{2}}\sigma_{(+)}
\otimes 1 , \nonumber\\
\chi &=& \sigma_3 \otimes \sigma_{(-)} \, ,\qquad \, \bar\chi =\sigma_3 \otimes
\sigma_{(+)} \, , \qquad \,
\xi = \sigma_3 \otimes 1 \, ,
\end{eqnarray}
where $\sigma_{\pm}=\frac{\sigma_1 \pm i \sigma_2}{2}$,  $\sigma_1$,
$\sigma_2$ and $\sigma_3$ are the Pauli matrices.

\section{The wave function for the zero energy state}

In the quantum theory the first-class constraints $H=0, S=0, \bar S=0$ and 
${\cal F}=0$ associated with the invariant action (\ref{action1}) under the
local $n=2$  conformal supersymmetry become conditions on the wave function 
$\Psi$. So that any
physical state must obey the following quantum constrains

\begin{equation}
\hat H \, \Psi =0, \hat S \, \Psi =0, {\hat{\bar{S}}}\, \Psi = 0 \,\, 
{\rm and} \,\,
\hat{\cal F} \, \Psi =0 ,
\label{physical}
\end{equation}
where the first equation in (\ref{physical}) is the Wheeler-DeWitt equation 
for the minisuperspace model. The eigenstates of the Hamiltonian (\ref{hamil}) 
have four-components 

\begin{equation}
\Psi (R,\varphi ) =
\left[ \begin{array}{c}
 \psi_1(R,\varphi ) \\
 \psi_2(R,\varphi ) \\
 \psi_3(R,\varphi ) \\
 \psi_4(R,\varphi )  
\end{array}
\right] \, .
\end{equation}

We have rewritten the equations $S\, \Psi=0$ and $\bar S \, \Psi=0$ in the 
following form 

\begin{eqnarray}
\left(\lambda\bar S - \bar\lambda S\right) |\Psi>& =& \left[- \left( 
\frac{A-A^\dagger}{2}\right) \left\{ \bar \lambda , \lambda \right\}
-\left( \frac{A+A^\dagger}{2} \right) \left[\bar\lambda ,\lambda \right]
\right. \nonumber \\
&&- \left. \left(\frac{B-B^\dagger}{2}\right)
(\bar\lambda\chi +\lambda\bar\chi )+\left( \frac{B+B^\dagger}{2} \right) 
(\lambda\bar\chi -\bar\lambda\chi )\right] |\Psi> =0, \\ 
\left( \bar\chi S -\chi\bar S\right)|\Psi> &=&\left[ \left(
\frac{A-A^\dagger}{2}\right)(\bar\chi\lambda +\chi\bar\lambda )
+\left(\frac{A+A^+}{2}\right)(\bar\chi\lambda -\chi\bar\lambda) \right. 
\nonumber \\
&& -\left. \left(\frac{B-B^\dagger}{2}\right)\left \{\bar \chi , \chi \right\}
+\left(\frac{B+B^\dagger}{2}\right) \left[\bar\chi ,\chi\right] 
\right]|\Psi> = 0 \, .
\end{eqnarray}

Using a matrix representation for $\lambda, \bar \lambda, \chi$ and
$\bar \chi$, we found that only $\psi_4$ have the right behaviour when
 $R\to \infty$ because $\psi_4 \to 0$, 
and due that the others components $\psi_1 , \psi_2$ and $\psi_3$ for the wave 
function at $R\to \infty$ are infinite, we consider these components as no
physical. Thus, there is a normalizable component $\psi_4$ for H such that 
$S \Psi = \bar S \Psi =0$, and this eigenstate corresponds to the
ground state with eigenvalue $E=0$. The wave function $\Psi$ has  also
the non-normalizable components  $\psi_1, \psi_2$ and $\psi_3$  for $E=0$,
but for them the eigenvalue of $H$ is non-zero, for this case all components
are normalizable

So, the partial differencial equations for $\psi_4 (R,\varphi)$ have 
the following forms

\begin{eqnarray}
\left[-R^{-\frac{1}{2}} \frac{\partial}{\partial R} - 6 g(\varphi)
R^{3/2}+
6\sqrt{k} M^2_{pl} R^{\frac{1}{2}}+\frac{3}{4} R^{-3/2} \right] \psi_4 &=&0,
\label{factor}\\
\left[\frac{\partial}{\partial \varphi}+2R^3 
\frac{\partial g(\varphi)}{\partial\varphi} \right] \psi_4 &=& 0 \, .
\label{scalar}
\end{eqnarray}

We get as solution for (\ref{factor}) and (\ref{scalar})

\begin{equation}
\Psi\equiv \psi_4 (R,\varphi)=C_o\,R^{3/4} e^{(-2g(\varphi)\,R^3+
3\sqrt{k} M^2_{pl} R^2)} \, .
\label{solution}
\end{equation}

The scalar product  for the solution (\ref{solution}) is normalizable 
with the measure $R^{\frac{1}{2}}\, dR \,d\varphi$ and for the superpotential 
$g(\varphi \to\pm\infty )\to \infty$. 

Consequently the solution 
(\ref{solution}) is the eigenstate of the Hamiltonian (\ref{hamil}) with 
 zero energy  and also with zero fermionic number.

The superpotential for the fluctuating scalar fields 
$\varphi =\varphi_0 + \tilde\varphi$ near the minimum of the potential 
$V(\varphi_0)=0$ has the following form  
(see (\ref{uno},\ref{dos},\ref{tres}, \ref{mass})).

\begin{eqnarray}
g(\tilde\varphi)&=&g(\varphi_0)+ 
\frac{\partial g(\varphi_0)}{\partial\varphi} \tilde\varphi 
+\frac{1}{2} \frac{\partial^2 g(\varphi_0)}{\partial\varphi^2}\tilde\varphi^2 
 \nonumber\\
&=& m_{3/2} M^2_{pl}\left[1 + \sqrt{\frac{3}{2}} \frac{\tilde \varphi}{M_{pl}}+
\frac{3}{4} \,\frac{\tilde \varphi^2}{M^2_{pl}}  \right]=
m_{3/2} M^2_{pl} \, f(x) \, ,
\end{eqnarray}
where $f(x)=1 + \sqrt{\frac{3}{2}} x+ \frac{3}{4} \,x^2$, with 
$x=\frac{\tilde \varphi}{M_{pl}}$.

So, as an example we can consider the case $k=0$ 
\begin{eqnarray}
 1&=& C_o^2 \int^{\infty}_{-\infty}
\int^\infty_0 R^{\frac{3}{2}} e^{-4g( \tilde\varphi)R^3} \,
R^{\frac{1}{2}}\, dR \,d\tilde\varphi \nonumber\\
&=&\frac{C_o^2}{12} \int^{\infty}_{-\infty}
\frac{d\tilde\varphi}{g(\tilde\varphi)}= \frac{C_o^2}{12 m_{3/2} M_{pl}}
\int^{\infty}_{-\infty} \frac{d x}{f(x)} = 
\frac{C_o^2 \sqrt {2} \pi}{12 \sqrt{3} m_{3/2} M_{pl}} \, ,
\end{eqnarray}
thus, the normalization constant has the following value
\begin{equation}
C_o=  \left( \frac{3}{2} \right)^{\frac{1}{4}} 
\sqrt{\frac{6 m_{3/2}\,M_{pl}}{\pi}}.
\end{equation}
The behaviour for the wave-function $\Psi$ in the $k=0$ case is shown in  
Figure 1.

The spectation value for the scalar factor $R$ with the chosen measure is

\begin{eqnarray}
\overline R &=&
<\Psi|R|\Psi> = C_o^2 \int^\infty_{-\infty} d \tilde\varphi \int^\infty_{0}
\left[ R^3 e^{-4g(\tilde\varphi)\, R^3} dR  \right] \nonumber\\
&=&\left( \frac{ \sqrt{3}\Gamma\left( \frac{4}{3}\right)}{4\pi 2^{1/6} } 
\int^\infty_{-\infty} \frac{dx}{\left[ f(x) \right]^{4/3}} \right)
\left( \frac{M_{pl}}{m_{3/2}} \right)^{\frac{1}{3}} \frac{1}{M_{pl}}\, ,
\end{eqnarray}
where $\Gamma(\frac{4}{3})$ is the Gamma function.

The size of the universe in the supersymmetry breaking state is of the 
order of
\begin{equation}
\overline R = C_1 \left(\frac{M_{pl}}{m_{3/2}} \right)^{\frac{1}{3}} 
\ell_{pl}\, ,
\end{equation}
where $\ell_{pl}= \frac{1}{M_{pl}}=\sqrt{8 \pi G_N}$ is the Planck length,
and $C_1$ has the following value
\begin{equation}
C_1= \frac{2^{1/6} \sqrt{3} \left(\frac{3}{4} \right)^{\frac{1}{3}}}{2 
\left( \frac{3}{8}\right)^{\frac{5}{6}}} 
\frac{\Gamma\left(\frac{4}{3} \right) 
\left[\Gamma\left(\frac{2}{3} \right)\right]^2 }{
\Gamma\left(\frac{5}{3}\right) 
\left[\Gamma\left(\frac{1}{3} \right)\right]^2   } \approx 0.505468 \, .
\end{equation}

\section{Conclusion}
Using the ideas of the supersymmetry breaking selection rules under local n=2 
conformal supersymmetry, we have presented  a solution to the wave function
$\Psi$ of the universe for the FRW cosmological model $(k=0)$, that is 
normalizable. For this proposal it was necessary to define one ``weighted'' 
inner product with the factor $R^{1/2}$ and  we constructed the corresponding
hermitian operators too.
Therefore, it  was  straighforward  to find the behaviour of the
 spectation value 
for the scalar factor like 
$\overline R = C_1 \left(M_{pl}/m_{3/2} \right)^{1/3} 
\ell_{pl}$, giving us the size of our universe.

In the proposed framework it is also possible to include the potential of 
hybrid inflation scenario \cite{LS,KS}, since the corresponding potential term 
can be easily introduced in the supersymmetric case. This subject will be 
reported elsewhere.

\section{\bf Acknowledgments}
Thanks to I. Lyanzuridi, E. Ivanov, S. Krivonos, 
L. Marsheva and A. Pashnev for their interest in this paper. 
This work was supported in part by CONACyT grant 3898P-E9608.

\newpage

\vskip 2ex
\centerline{
\epsfxsize=280pt
\epsfbox{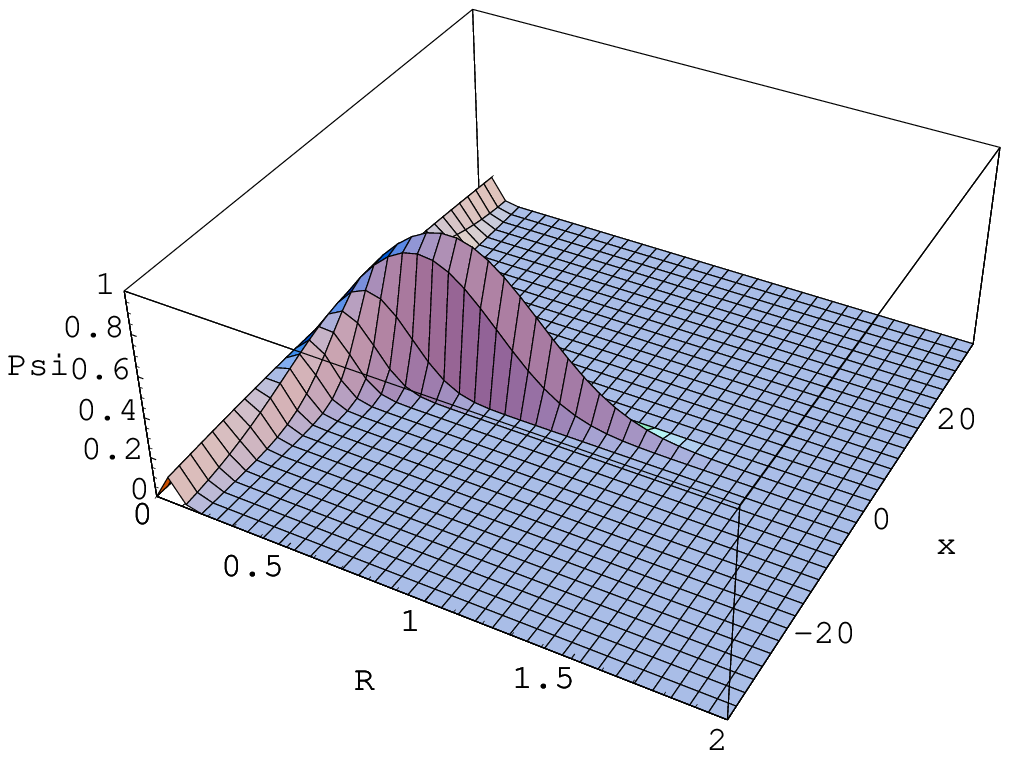}}
\vskip 4ex
\begin{center}
{\small{Fig. 1}\\
The more detailed plot of the region of the definition $\{ R, x \} \to
\{[0,\infty ], [-\infty,\infty ]  \}$ for the normalized wave-function $\Psi$
for k=0 
case. Here it can be seen the asimptotic behaviour when 
$x=\tilde \varphi/M_{pl} \to \pm \infty$. In this plot for simplicity we have
taked units in where $m_{3/2}=M_{pl}=1$.
}
\end{center}

\end{document}